\documentclass[a4paper]{llncs}  	
\usepackage{geometry}                		
\usepackage{graphicx}				
\usepackage{amssymb}

\usepackage{amsmath}

\usepackage{graphicx}

\usepackage{rotating}

\usepackage{prooftree}

\usepackage{url}
\newcommand{\keywords}[1]{\par\addvspace\baselineskip
\noindent\keywordname\enspace\ignorespaces#1}

\usepackage{z-eves}

\usepackage{xargs}

\usepackage[pdftex,dvipsnames,table]{xcolor}  

\usepackage{subfig}

\newcommand{\tactic}[1]{\textsf{#1}}
\newcommand*{\timesfont}{\fontfamily{ptm}\selectfont}
\DeclareTextFontCommand{\texttimes}{\timesfont}

\usepackage[bordercolor=gray!20,backgroundcolor=blue!10,linecolor=none,textsize=footnotesize,textwidth=1.2in]{todonotes}
\reversemarginpar
\newcommandx{\SR}[2][1=]{\todo[linecolor=Plum,backgroundcolor=Plum!25,bordercolor=Plum,#1]{Reeves: #2}}
\newcommandx{\PR}[2][1=]{\todo[linecolor=Goldenrod,backgroundcolor=Goldenrod!25,bordercolor=Goldenrod,#1]{Rimba: #2}}
\newcommandx{\thiswillnotshow}[2][1=]{\todo[disable,#1]{#2}}

\newcommand{\ie}{\emph{i.e.}}
\newcommand{\etc}{\emph{etc.}}
\newcommand{\eg}{\emph{e.g.}}

\title{Usable-by-Construction---a formal framework}
\author{Steve Reeves}
\institute{Department of Computer Science, University of Waikato}

\begin{document}

\begin{titlepage}

\maketitle

\end{titlepage}

{\bf NOTE:} This paper completes the work started in a workshop paper at EICS in 2019 \cite{TrendsEICS19} by giving a complete set of tactics and logical rules and also shows a slightly larger example in action, as well as extending the discussion further in various places.

\begin{abstract}
We propose here to look at how abstract a model of a usable system can be, but still say something useful and interesting, so this paper is an exercise in abstraction and formalisation, with usability-of-design as an example target use.

We take the view that when we claim to be designing a usable system we have, at the very least, to give assurances about its usability properties. This is a very abstract notion, but provides the basis for future work, and shows, even at this  level that there are things to say about the (very concrete) business of designing and building usable, interactive systems.

Various forms of verification and validation can provide a high level of assurance but it can be very costly, and there is clearly a lot of resistance to doing things this way. 

In this paper, we introduce the idea of usable-by-construction, which adopts and applies the ideas of correct-by-construction to (very abstractly) thinking about usable systems. 

We give a set of construction rules or tactics to develop designs of usable systems, and we also formalize them into a state suitable for, for example, a proof assistant to check claims made for the system as designed. 

In the future, these tactics would allow us to create systems that have the required usability properties and thus provide a basis to a usable-by-construction system. Also, we should then go on to show that the tactics preserve properties by using an example system with industrial strength requirements. And we might also consider future research directions.
\\

{\bf Research highlights: }
\begin{itemize}
\item
Abstraction is a key design method;
\item
Compositionality helps to structure designs
\end{itemize}

\end{abstract}

\section{Introduction}

The position taken in this paper is that more abstraction in the methods for modelling and designing interactive systems would be a good thing, certainly as a starting point for design. The paper can be seen as an exercise in formalisation and also pushing abstraction to an extreme. 


The aim of this paper is to introduce the idea of \emph{usable-by-construction}. This, in essence, takes the ideas of \emph{correct-by-construction} that formed much of the work of Dijkstra \cite{Dij76}, Gries \cite{Gri89} and many others, and applies them to the problem of usable systems. In particular we want to see how we can develop a set of construction rules or tactics which allow us to build designs of usable systems without having to perform, say, \emph{post hoc} verification on the constructed system. That is, we want tactics that can build \emph{only} usable systems: any system built with the tactics \emph{will necessarily} have the required usabilty properties simply due to the nature of the construction tactics themselves.

Since we are trying to hit the spot between maximum abstraction and maximum simplicity, we leave the question ``what exactly do you mean by usable?'' unanswered. If pressed for an answer we would say that our abstraction allows \emph{any} answer to that question that you personally are happy to accept. Here we are totally agnostic about what usable means. Think of the definition of usable as being a parameter of our rules.

%
%
%

\subsection{Related work}

Within the formal methods for HCI world there are no directly related works to what we are presenting here. This is not surprising since we have gone to extremes to be abstract. However, there is a very long and distinguished seam of work recorded in the DSV-IS series (1994-2008) of conferences \cite{DSV-IS}, in the FMIS workshop series (for example \cite{FMIS07},\cite{FMIS13}),  and, more recently (and continuing the DSV-IS tradition) the EICS series of conferences \cite{EICS} and the (now) associated ACM journal \cite{PACM}. There is also the work of Celentano and Dubois \cite{CeD15} on metaphors, which give one way of seeing the range of possibilities for instantiating our abstract models. There is also the work of Bowen and Reeves \cite{BoR07a} and their various papers. Both these bodies of work also consider the role of refinement or reification which, as we note later, is a needed step towards instantiation and (ultimately, if required) implementation. In the case of Bowen and Reeves \cite{BoR07a} the formality introduced here can be preserved through the subsequent instantiation and implementation to a high degree, and Celentano and Dubois's \cite{CeD15} work, while not allowing preservation of formality, does show how informal reification can be used instead.

All that said, the work in this paper is entirely prior to all of that referred to and cited above.

\subsection{Summary}

In this paper we are doing two things: developing a framework of rules for building a usable system; ensuring that rules we define enforce certain good design and development techniques. 

The aims of these techniques can be summarised by saying that we are trying to bring some good engineering principles to bear, namely:

\begin{itemize}
\item
model a system before going to the expense of building it;
\item
use maths to check that the system is fit for purpose;
\item
for a good design, don't get concrete too early or we'll lock ourselves into design choices too soon;
\item
build a system, not by adding features at will and on the fly---but in a controlled, structured way as this keeps complexity under control.
\end{itemize}

These points are usually summarised as---
\begin{itemize}
\item
modelling
\item
maths 
\item
abstraction 
\item
compositionality (i.e. start with a small set of very simple, basic actions and then have a few simple rules which given already constructed pieces allows us to compose them into new, larger systems).
\end{itemize}

\section{Basic definitions}

We assume that each system is made up of components (which might be people, computers, software systems and so on right down to simple widgets) and connections between them. A connection represents a use of one component by the other: if $c$ uses $d$ then we have a connection $\langle c, d \rangle$. That is almost all we say, so we are going for maximum generality here. 


We make the following definitions. 

\begin{definition}
A \emph{system} is $\langle C,N\rangle$, where $C$ is a component set $\{{c_1,c_2,...,c_n}\}$, and $N$ is a connection set, $\{{n_1,n_2,...,n_m\}}$, where each $n_j = \langle c_i,c_k\rangle$ for some $c_i,c_k \in C$.
\end{definition}

This definition is a starting point, but it clearly too general to be very interesting. With our problem in mind, namely designing usable systems, we introduce two further ideas. Firstly, we have a subset of $C$, called $I$, which is a set of components which can be interacted with. Secondly, each interactive component, $i \in I$, is associated with its own set of components that are allowed to interact with it, $A_i$, which we refer to as \emph{an interactive component's allowed set}. 

For a component to be allowed access to an interactive component, that component needs to be added into the allowed set of that interactive component. These sets of ``allowed" components can be thought of as expressing propositions about the system.

The key point here is: assuming that allowing the components access to the interactive components is sensible or allowable (a decision of the designers, perhaps based on experimental or past design knowledge or experience, \etc), the system as whole is accepted as usable. 

The sets can be thought of sets of permissions too: if a component is in an interactive component's allowed set then that component has permission to use that interactive component while keeping the whole system ajudged as usable. 


 We refer to systems with these two additional sets as \emph{acceptable systems}.

\begin{definition}
An \emph{acceptable system} $\langle C,N,I,A\rangle$ extends a system $\langle C,N\rangle$, where $I \subseteq C$ is a set of components that are interactive and $A$ is a family of sets of components $\bigcup_i A_i$, where for $i \in I$, $A_i \subseteq C$. $A_i$ is a set of components allowed (for agreed reasons) to use the interactive component $i$.
\end{definition}

Now we need to think of how we can combine such systems, via \emph{tactics}, preserving usability---so we are going for a compositional approach here. These tactics are \tactic{connect}, \tactic{disconnect}, \tactic{create}, and \tactic{delete}. 


In what follows we make the (surely benign) assumption that every component has access to itself.

\section{Tactics}

We introduce various rules (which we call tactics) for building systems from smaller systems, and ultimately from single components, with the aim that the result of each tactic, assuming we start with usable systems, will necessarily be usable systems. Note that this system is very liberal, in the sense that connections between systems containing interactive components  are generally allowed, but we also without exception keep track of what components have access to what interactive components. So, usability here comes about because we are completely open and honest about who has access to what. 

Later, as we will see, we make the modelling more \emph{realistic} by introducing the notion of ``usability-enhancing" (to be discussed) and we use this to control which components have access to interactive components. 

Recall that our overall plan is to have simple rules to start with, see how far we can get, then introduce further rules carefully to get us closer to our aim, without disturbing (as far as possible) the simplicity of the modelling.

To help the presentation we introduce the notion of a path:

\begin{definition}Given a system $\Sigma = \langle C, N, I, A \rangle$,  a path exists between components $c$ and $c'$ in $C$, which we write as $c \rightsquigarrow_{\Sigma} c'$, iff either:
\begin{enumerate}
\item
$c$ is connected to $c'$, i.e. $\langle c, c' \rangle \in N$, or
\item
there is some $d \in C$ such that $c \rightsquigarrow_{\Sigma} d$ and $\langle d,c' \rangle \in N$
\end{enumerate}
\end{definition}
We drop the subscript where the context allows, and we iterate the notion of path, so that, for example, $c \rightsquigarrow d \rightsquigarrow e$ abbreviates $c \rightsquigarrow d \land d \rightsquigarrow e$. Also, by $\forall c \in a \rightsquigarrow b.P$ we mean that all components on the path between $a$ and $b$ satisfy the predicate $P$.

\subsection{Connect Tactic}
\label{SEC:Connect}
The \tactic{connect} tactic adds a connection between two components of a system under certain conditions. (We can also see this rule as allowing us to join two such systems together via the joined components. In the full treatment we have rules to deal with this eliding of meaning.)
This is done by creating a connection between two components. 

\begin{definition}Connects to\newline
\label{def:connect}
If $\Sigma = \langle C,N, I,A\rangle$ 
then making a new connection between $c_a$ and $c_b$ in $C$ means creating $\Sigma' = \langle C, N  \cup \{ \langle c_a,c_b\rangle \}, I, A \rangle,$ and $\langle c_a,c_b\rangle \notin N$ with the condition that: 
\begin{center}
$\forall i \in I. \forall c, d' \in C. \forall d \in c_b \rightsquigarrow d'. \forall c' \in c \rightsquigarrow c_a. d \in A_i \implies   c' \in A_i$
\end{center}
\end{definition}

The condition here simply says that any elements on any path  $c \rightsquigarrow c_a$  that gets joined to a path $c_b \rightsquigarrow d$, where this path contains elements allowed access to any interactive component, must already be allowed access to those same interactive components. This is, of course, a very general condition, i.e. it means that almost every component in any system might have to be allowed access to almost all interactive elements in that system. For the moment, though, we are concerned only with giving rules which preserve usability. 

Later, when we introduce the idea of usability-enhancing components, we will see that these (very large) sets of (general) components that are granted access to interactive components can be made far smaller.

Some simple results follow directly from this definition. For example, given some system containing $c$ and $i$, if $c \rightsquigarrow i$ for $i$ an interactive component, then $c \in A_i$, which is itself a special case of the more general result that if $c \rightsquigarrow d$ and $d \in A_i$ for some interactive component $i$, then $c \in A_i$. 

\subsection{Disconnect Tactic}
\label{SEC:Disconnect}
The \tactic{disconnect} tactic removes a connection between a source component and a target component. This is done by removing a connection between two components in the connection graph. Structurally, a use of the target component is revoked from the source component. (It may be that this gives us two completely disconnected sets of components.) 

The \tactic{disconnect} tactic requires two parameters, i.e. the source component and the target component. 

\begin{definition}Disconnects from\\
\label{def:disconnect}
If $\Sigma= \langle C,N,I,A\rangle$, then disconnecting $c_a \in C$ from $c_b \in C$ means deleting a connection, $\langle c_a,c_b\rangle \in N$, and creating $\Sigma' = \langle C, N',I, A\rangle$, where $N'= N\ \setminus \langle c_a,c_b\rangle$
\end{definition}

\subsection{Create Tactic}
\label{SEC:Create}
The \tactic{create} tactic creates a new component in the system. This is done by creating a new component in the connection graph. 

The \tactic{create} tactic requires two parameters, \ie\ the component name and the component type (\ie\ general or interactive). Creating an interactive  component requires an additional piece of information: a set of components that are allowed to have access to that component. Note that here we use an \emph{override} operator $\oplus$: this updates a function at a point in its domain so that it has a new value at that point.

\begin{definition}Creates\\
\label{def:create}
If $\Sigma = \langle C,N,I,A\rangle$, then creating a new general (\ie\ non-interactive) component, $c \notin C \wedge c \notin I $, means creating $\Sigma' = \langle C \cup \{c\},N,I,A\rangle$. Creating a new interactive component, $i \notin C \wedge i \notin I$, means creating $\Sigma' = \langle C \cup \{i\},N,I\cup \{i\},A \oplus \{i \mapsto A_i \cup i\}\}\rangle$.
\end{definition}

Definition \ref{def:create} covers the case for creating the first component for a system. When creating the first component we start with $\Sigma = \langle \{\},\{\},\{\},\{\}\rangle$. After the creation of that first component $c$, the system will then be $\Sigma' = \langle \{c\},N,I,A\rangle$, where $c$ may be in $I$ if it is interactive.


\subsection{Delete Tactic}
\label{SEC:Delete}
The \tactic{delete} tactic removes a component in the system. This is done by deleting a component in the connection graph. However, the component needs to be isolated before the \tactic{delete} tactic is used. A component is isolated when it has no connections to other components in the system. 

The \tactic{delete} tactic only requires one parameter, \ie\ the component to be removed.
\begin{definition}Deletes\\
\label{def:delete}
If $\Sigma = \langle C,N,I,A\rangle$, then deleting a non-interactive $c \in C \wedge c \notin I$, means creating $\Sigma' = \langle C \setminus \{c\},N, I, A \rangle$. Deleting an interactive component, $i \in I$, means creating $\Sigma' = \langle C \setminus \{i\},N, I \setminus \{i\},A \setminus A_i\rangle$. Delete is allowed iff the component is isolated, \ie\ concerning $c$,  $\forall c_i \in C \setminus \{c\} \cdot \langle c,c_i\rangle \notin N \wedge \forall c_i \in C \setminus \{c\} \cdot \langle c_i,c\rangle \notin N$
\end{definition}


\subsection{Allow Tactic}
\label{SEC:Allow}

The \tactic{allow} tactic allows a component to have access to an interactive component. This is done by adding a component to the set of components that are allowed to have access to the interactive one.

\begin{definition}Add to Allowed set\\
\label{def:grant}
If $\Sigma = \langle C,N,I,A\rangle$, then allowing $c_i$ access to $c_j$ means creating $\Sigma' = \langle C,N,I,A'\rangle$, where $A'= A \oplus \{c_j \mapsto A_{c_j} \cup \{c_i\}\}$ iff $c_j \in I.$ Otherwise, $A' = A$. 
\end{definition}

The \tactic{allow} tactic loosens the restriction of the \tactic{connect} tactic. Given this loosening, we have to be careful about what we claim for a system constructed with our tactics. In particular, we have to ensure that the \emph{assumption} that the family of sets of components $A$ have been allowed access to interactive components is made explicit in any \emph{guarantees} we give about the system constructed. So, if we have constructed the system $\langle C,N,I,A\rangle$ then we have to say that:

\begin{quote}
$assuming$ that the family of sets of components $A$ have correctly been allowed to access certain interactive components, \emph{then we guarantee} that the constructed system is usable
\end{quote}

which we might write formally as\footnote{The symbol $\vDash$ is borrowed from formal logic, and there it is usually called a turnstile. These are conventionally used to seperate assumptions from conclusions, hence our use of the symbol here}:
\[
A \vDash <C,N,I>
\]
So, when we ``hand" a system to a client, we hand them something that, as long as they use it in the right context, \ie\ $any$ context in which the assumption is satisfied, \ie\ $any$ context where it is permitted to allow the interactions that have been allowed to the components that they have been granted to, then we guarantee that the system is usable. Stated alternatively (as we mentioned above in a previous section) we can think of $A$ as recording the permissions for accessing interactive components, so $A \vDash <C,N,I>$ is saying that assuming that we are happy to allow the permissions as given in $A$, then the system as described by $<C,N,I>$ is usable. 

It is useful to think of this notation as stating a \emph{contract} between modeller and client. It makes plain exactly what is being assumed ($A$), and exactly what may then be taken to be a usable system ($<C,N,I>$) \emph{under those assumptions}.

There are, in particular, no conditions put on what sort of context we are talking about here; all we are doing is saying that \emph{as long as the context of use can satisfy the assumptions stated} then this is a usable design. We are recording assumptions relating to the structure of the recorded design, and terms like usable, interactive component etc. are left up to the user to instantiate as they see fit. They are like parameters to the design and, as ever, a conversation between whoever builds the system according a design given in the manner we are showing here and the end-user needs to be had in order that the correct instantiations of these terms is made.

\subsection{Revoke Tactic}
\label{SEC:Revoke}

The \tactic{revoke} tactic disallows a component access to an interactive component. This is done by removing a component from the set components that are allowed to have access to that interactive component.  We have to restrict use of this tactic to situations where the component whose access is being revoked is not on any path that contains the interactive component concerned.

\begin{definition}Revoke from Allowed set\\
\label{def:revoke}
If $\Sigma = \langle C,N,I,A\rangle$, then revoking $c$'s access to interactive component $i$ means creating $\Sigma' = \langle C,N,I,A'\rangle$, where $A'= A \oplus \{i \mapsto \{A_{i} \setminus \{c\}\}$ with the restriction that there is no path that both $c$ and $i$ are on, i.e. it is not that case that $c \rightsquigarrow_{\Sigma} i$.
\end{definition} 



\section{The rules}

In this section we re-state the rules above in a more formal setting. This does two things: it makes clear exactly what is being assumed and what is being concluded; and it allows us to move towards a \emph{logic} for usable systems, which itself (via a proof assistant, theorem-prover or other programmed form of the rules coupled with some search strategy) leads to algorithmic construction of usable systems. We expand on these points as follows: 

\begin{itemize}
\item
they allow goal-directed construction (examples below), because a rule read backwards (or upwards) tells us what we must show in order to have the conclusion we desire;
\item
they are good for design in general since such rules---
\begin{itemize}
\item
provide some guidance (the shape of the desired system determines, to some extent, the rules that must be used to build it);
\item
suggest a pattern to look for (the use of a restricted set of rules soon gives rise to repeated patterns of development, which then gives rise in turn to derived rules which usefully encode recurring, common patterns);
\item
ease explanation (the structure suggest the form and content of answers to the question: how was this system constructed, and why is it usable?);
\item
promote understanding (see: all the above);
\end{itemize}
\item
although we trade away complete flexibility (\ie\  on the fly, {\it ad hoc} design), we gain better understanding, structure, robustness \etc ;
\item
they take us towards a method for checking and building systems: the rules, being formal, can easily be read as algorithms.
\end{itemize}

The rules will (following standard methods) follow from the definitions of the tactics that we have given earlier in the paper. 
Note that all of the tactics essentially tell us how to add to or delete from sets of values of various sorts (components, connections, certain subsets of components, along with certain conditions \etc) and these sorts of definitions, while defining the vocabulary of a way of developing our systems, do not extend the logical basis that we are working from (which is just standard set theory for us here). We say that our system is a conservative extension of set theory: conservative because it does not introduce any new rules to the underlying set theory: we might say that, via the new terms introduced via the definitions, \ie\ the new vocabulary, we ``repackage" combinations of existing set theory rules to give us rules in the vocabulary of our usable systems. 

So, the logic of set theory can be used to express, via our definitions, a logic for our system. To give a flavour of this, consider the following (artificially simple) definition:
$$
S =_{def} \{x \in T | P(x) \}
$$
where $T$ is some already known set and $P$ is some predicate over members of $T$. This defines the new set $S$ in terms of $T$ and $P$.

From this definition we can read-off rules which allow us to introduce sets like $S$, and to deconstruct them too.

So we have rules like:

\begin{minipage}{0.3\linewidth}
\[
\begin{prooftree}
x \in T \qquad
P(x)
\justifies
x \in S
\using
{S^+}
\end{prooftree}
\]
\end{minipage}%
\hfill
\begin{minipage}{0.3\linewidth}
\[
\begin{prooftree}
x \in S
\justifies
x \in T
\using
{S^-_1}
\end{prooftree}
\]
\end{minipage}%
\hfill
\begin{minipage}{0.3\linewidth}
\[
\begin{prooftree}
x \in S
\justifies
P(x)
\using
{S^-_2}
\end{prooftree}
\]
\end{minipage}
\vspace{.5cm}

%

\subsection{Creates and Deletes}

Creates is easy, and has two introduction rules, one for each class of component (general or interactive).

First, a general, non-interactive component is created:
$$
\begin{prooftree}
A  \vDash  \langle C, N, I \rangle  \qquad c \notin C 
\justifies
A \vDash  \langle C \cup \{c\}, N, I \rangle
\using
{create^+_1}
\end{prooftree}
$$

Then, an interactive component:

$$
\begin{prooftree}
A  \vDash  \langle C, N,  I \rangle  \qquad i \notin C 
\justifies
A \oplus \{i \mapsto \{i\}\} \vDash  \langle C \cup \{i\}, N, I \cup\{i\} \rangle
\using
{create^+_2}
\end{prooftree}
$$

Deletes is slightly more complicated since we have to make sure that the component we are deleting is isolated, which means that there are no connections either into or out of it and that it appears in no interactive component's allowed set.

These two conditions are enforced in the premises of the rules: the first by requiring that in order to delete component $c$ we have
$$
\forall c_i \in C \setminus \{c\} \cdot \langle c,c_i\rangle \notin N \wedge \forall c_i \in C\setminus \{c\} \cdot \langle c_i,c\rangle \notin N
$$
which we denote by $isolated~c$, and the second by requiring that for a system with interactive set $I$ and assumption set $A_i$ for $i \in I$, where we are deleting component $c$
 $$
 \forall i \in I. c \notin A_i
 $$
 
The first rule talks about deleting a non-interactive component:
$$
\begin{prooftree}
A  \vDash  \langle C, N,  I \rangle  \qquad c \in C \qquad c \notin I \qquad isolated~c \qquad  \forall i \in I. c \notin A_i
\justifies
A \vDash  \langle C \setminus \{c\}, N,  I \rangle
\using
{delete^+_1}
\end{prooftree}
$$

Then, we have a rule to delete an interactive component:

$$
\begin{prooftree}
A  \vDash  \langle C, N, I \rangle  \qquad i \in C \qquad i \in I \qquad isolated~i \qquad  \forall i' \in I. i \notin A_i'
\justifies
A \setminus A_i \vDash  \langle C \setminus \{i\}, N,  I  \setminus \{i\} \rangle
\using
{delete^+_2}
\end{prooftree}
$$

and note that we have allowed interactive components to be allowed access to other such components.

\subsection{Disconnect rule}
Consider the disconnect tactic, and recall its definition:

\begin{quote}
{\it Disconnects from\\
If $\Sigma= \langle C,N,I,A\rangle$, then disconnecting $c_a \in C$ from $c_b \in C$ means deleting a connection, $\langle c_a,c_b\rangle \in N$, and creating $\Sigma' = \langle C, N',I, A\rangle$, where $N'= N\ \setminus \langle c_a,c_b\rangle$}
\end{quote}

This gives us two rules: one ``introduction" rule for moving from a system with a certain connection to one where a disconnection (\ie\ removal of that certain connection) has happened:
$$
\begin{prooftree}
A \vDash \langle C,N,I\rangle \qquad \langle c_a, c_b\rangle \in N
\justifies
A \vDash \langle C,N \setminus \langle c_a,c_b\rangle,I\rangle
\using
{disconnect^+}
\end{prooftree}
$$
and an ``elimination" rule which, given a system that has had a disconnection performed on it, can ``reverse" this (somewhat artificially perhaps, but it is a rule we gain nonetheless):
$$
\begin{prooftree}
A \vDash \langle C,N,I\rangle \qquad c_a, c_b \in C
\justifies
A \vDash \langle C,N \cup \{ \langle c_a,c_b\rangle \},I\rangle
\using
{disconnect^-}
\end{prooftree}
$$
%
%
\subsection{Connect rule}

Recall that we assume that all components are connected to themselves.
$$
\begin{prooftree}
A  \vDash \langle C, N, I\rangle \quad c_a \in C \quad c_b \in C \quad 
\[\forall i \in I. \forall c, d' \in C. \forall d \in c_b \rightsquigarrow d'. \forall c' \in c \rightsquigarrow c_a. d \in A_i \implies   c' \in A_i \leadsto
d \in A_i \implies   c' \in A_i\]
\justifies
A  \vDash \langle C, I \cup \langle c_a, c_b\rangle, I\rangle
\using
{connect^+}
\end{prooftree}
$$
Note that we may have to use $allowed^+$ (coming soon) before these rules in order that we can connect to an interactive component.
\subsection{Revoke and allowed}

Rules for taking away and adding something from the set of those things allowed access to some interactive component.

$$
\begin{prooftree}
\Sigma \qquad  c_j \in I \qquad c_i \in A_{c_j}  \qquad \langle c_i, c_j \rangle \notin N  \qquad c \in C, \langle c_i, c \rangle \in N, c_i \notin I \vdash c \notin C_{c_j} \lor c \in I
\justifies
A \oplus \{ c_j \mapsto A_{c_j} \setminus \{c_i\}\} \vDash  \langle C, N, I \rangle
\using
{revoke^+}
\end{prooftree}
$$

We could have  expressed this as two introduction rules, one for each of the two disjuncts in the final premise, \ie\ two different final premises:

$$
\begin{prooftree}
\Sigma \qquad  c_j \in I \qquad c_i \in A_{c_j}  \qquad \langle c_i, c_j \rangle \notin N \qquad c \in C, \langle c_i, c \rangle \in N, c_i \notin I \vdash c \notin C_{c_j} 
\justifies
A \oplus \{ c_j \mapsto A_{c_j} \setminus \{c_i\}\}\vDash  \langle C, N, I \rangle
\using
{revoke^+_1}
\end{prooftree}
$$

$$
\begin{prooftree}
\Sigma \qquad  c_j \in I \qquad c_i \in A_{c_j}  \qquad \langle c_i, c_j \rangle \notin N \qquad c \in C, \langle c_i, c \rangle \in N, c_i \notin I \vdash c \in I
\justifies
A \oplus \{ c_j \mapsto A_{c_j} \setminus \{c_i\}\} \vDash  \langle C, N, I \rangle
\using
{revoke^+_2}
\end{prooftree}
$$

(Note: each final premise in each of these rules uses an alternative, standard, notation for introducing assumptions that are local to a premise. It is a syntactic variant to the way this was done in the $connect^+$ rule above. This is our preferred notation; the previous version was used just for reasons of space.)

The point of the allowed set for some interactive component is that it records our actions in granting access since this forms part of our contract. Recall that $A \vDash \langle C, N,I \rangle$  simply means that assuming we have correctly (acceptably) granted access as recorded in $A$, then the system is usable. 
$$
\begin{prooftree}
A  \vDash  \langle C, N, I \rangle \qquad i \in I \qquad c \in C
\justifies
A \oplus \{ i \mapsto A_i \cup \{ c \}\} \vDash  \langle C, N, I \rangle
\using
{allowed^+}
\end{prooftree}
$$

\subsection{General rules}\label{SEC:genrules}

We will allow any first-order logic and set theory rules to be used within our proofs. The only other rules that are specific to our systems will be to do with naming conventions. One commonly used rule set (which makes the statement of some of the other rules more straightforward) allows us to switch between using a single designation for a system where, in fact, the components form disjoint sets (and therefore all the connections and interactive components are in disjoint sets too, and so they can be though of as two different systems). Going from a single system to one which is actually two disjoint systems can happen when we delete connections, for example. But sometimes when we delete a connection we stay with a single system (\ie\ the components don't fall into disjoint sets). Rather than have different rules for these different cases it is more convenient to allow a system which is actually two disjoint systems to be named as either one system or two. The following rules make all this clear:

$$
\begin{prooftree}
A^a  \vDash  \langle C^a, N^a, I^a \rangle  \qquad A^b  \vDash  \langle C^b, N^b, I^b \rangle \qquad C^a \cap C^b = \emptyset
\justifies
A^a \cup A^b \vDash  \langle C^a \cup C^b, N^a \cup N^b, I^a \cup I^b \rangle
\using
{naming_1}
\end{prooftree}
$$

$$
\begin{prooftree}
A \vDash  \langle C, N, I \rangle \quad C = C^a \cup D \quad C^a \cap D = \emptyset
\justifies
A^a  \vDash  \langle C^a, N^a, I^a \rangle 
\using
{naming_2}
\end{prooftree}
$$

The first rule here says that two disjoint systems a can be thought of as a single system. The second says that if we have any single system which actually consists of two disjoint systems then the two systems can be picked out. (Note: once $C^a$ is determined then $N^a$ and $I^a$ and $A^a$ can be calculated, and the same for $C^b$. Also, if we chose $D$ to be $\emptyset$ then $naming_2$ becomes trivial, since it says that a system can be replaced by itself, because the premises mean that $C = C^a$.) 

Finally we have an axiom (\ie\ a rule with no premises, which just says that its conclusion is proved without any further work) for the simplest (useless) system as mentioned before:

\begin{center}
\begin{prooftree}
\justifies
\emptyset  \vDash  \langle \emptyset, \emptyset, \emptyset \rangle 
\using
{axiom}
\end{prooftree}
\end{center}

\subsection{Tiny example}

\subsubsection{$connect^+$ and $disconnect^+$ are inverses:}

If we make a new connection and then disconnect immediately afterwards we get back to the same system that we started with. 

\begin{sidewaysfigure}

\begin{prooftree}
\[ 
A  \vDash \langle C, N, I\rangle \quad c_a \in C \quad c_b \in C \quad \[ \forall i \in I. \forall c, d' \in C. \forall d \in c_b \rightsquigarrow d'. \forall c' \in c \rightsquigarrow c_a \leadsto
d \in A_i \implies   c' \in A_i\]
\justifies
A  \vDash \langle C, N \cup \langle c_a, c_b \rangle, I\rangle \quad
\using{connect^+}
\]
\quad 
\[ 
\justifies
\langle c_a, c_b \rangle \in N \cup \langle c_a, c_b \rangle
\using{by\ set\ theory}
\]
\justifies
A  \vDash \langle C, N, I \rangle
\using
{disconnect^+}
\end{prooftree}
\caption{Proof that connecting followed by disconnecting leaves a system unchanged}
\label{fig:smallproof}


\vspace{.75cm}

{\tiny

\begin{prooftree}
 \[
  \[
   \[
    \[
\justifies
\emptyset \vDash \langle \emptyset, \emptyset, \emptyset \rangle
\using{axiom}
    \]
\quad
    \[
\justifies
A \notin \emptyset
\using{ST}
    \]
\justifies
\{A \mapsto \{A\}\} \vDash \langle \{A\},\emptyset, \{A\} \rangle
\using{create_2^+}
   \]
\quad
   \[
\justifies
G \notin \{A\}
\using{ST}
  \]
\justifies
\{A \mapsto \{A\}\} \vDash \langle \{A,G\}, \emptyset, \{A\} \rangle
\using{create_1^+}
 \]
\quad
 \[
\justifies
H \notin  \{A,G\}
\using{ST}
 \]
\justifies
\{A \mapsto \{A\}\} \vDash \langle \{A,G,H\}, \emptyset, \{A\} \rangle
\using{create_1^+}
\]
\quad
\[
\justifies
A \in \{A\}
\using{ST}
\]
\quad
\[
\justifies
G \in \{A,G,H\}
\using{ST}
\]
\justifies
\{A \mapsto \{A,G\}\}   \vDash \langle \{A,G,H\}, \emptyset, \{A\} \rangle
\using{granted^+}
\end{prooftree}
\caption{Fragment for proof in Figure \ref{fig:AGH}}
\label{fig:fragment 1}
\vspace{0.75cm}

\begin{prooftree}
  \[
Insert\ Figure\ \ref{fig:fragment 1}
\quad
   \[
\justifies
A \in \{A\}
\using{ST}
   \]
\quad
   \[
\justifies
H \in \{A,G,H\}
\using{ST}
   \]
\justifies
\{A \mapsto \{A,G,H\}\}   \vDash \langle \{A,G,H\},\emptyset, \{A\} \rangle 
\using{granted^+}
  \]
\quad
  \[
\justifies
A \in \{A,G,H\} 
\using{ST}
  \]
\quad 
  \[
\justifies
G \in \{A,G,H\} 
\using{ST}
  \]
  \quad
 \[
\justifies
A  \in \{A,G,H\} \implies G \in \{A,G,H\}
\using{ST}
 \]
\justifies
\{A \mapsto \{A,G,H\}\}   \vDash \langle \{A,G,H\}, \{\langle G, A\rangle\}, \{A\} \rangle 
\using{connect^+}
\end{prooftree}
\caption{Part of construction of system using $A$, $G$ and $H$ in Figure \ref{fig:AGH}}
\label{fig:fragment 2}
\vspace{0.75cm}

\begin{prooftree}
Insert\ Figure\ \ref{fig:fragment 2}
\quad
\[
\justifies
A \in \{A,G,H\} 
\using{ST}
\]
\quad 
\[
\justifies
H \in \{A,G,H\} 
\using{ST}
\]
\quad 
\[
\justifies
A,G  \in \{A,G,H\} \implies H \in \{A,G,H\}
\using{ST}
\]
\justifies
\{A \mapsto \{A,G,H\}\} \vDash \langle \{A,G,H\}, \{\langle G, A\rangle, \langle H, G \rangle\}, \{A\} \rangle
\using{connect^+}
\end{prooftree}
\caption{Construction of system using $A$, $G$ and $H$ in Figure \ref{fig:AGH}}
\label{fig:AGHproof}

}

\end{sidewaysfigure}

\begin{figure}
\centering
\includegraphics[height=5cm]{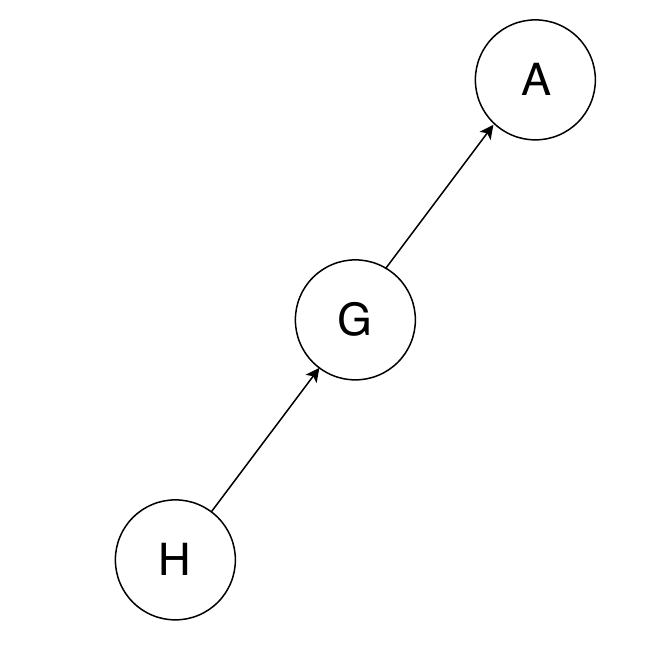}
\caption{A tiny system}
\label{fig:AGH}
\end{figure}

%
%
%
As we can see from this tiny proof in Figure \ref{fig:smallproof}, under the assumptions that we start with the system $A \vDash \langle C, N, I \rangle$ where $c_a \in C$ and $c_b \in C$  and assuming all the components involved respect the accessibility  requirements of $i$ as summarised in $A_i$ (which is what the fourth---big--premise is saying), which allow a connection to happen, then undoing the connection results in the system $A \vDash \langle C, N, I \rangle$ we started with.

This is about the simplest general property we would expect to be provable if the rules have correctly captured the intended meaning of such systems and their properties. Showing that such proofs are possible is part of the usual validation process for any formalisation.

 
We also give the proof required to construct a small part of a system, where we prove that the fragment where  $G$ and $H$ connect with interactive component $A$, as in Figure \ref{fig:AGH}, is constructible.

\section{Usage}

Apart from giving us a logic for reasoning about systems, these rules can help guide us in the construction of systems. For example, say we had to construct a system of the form $A \vDash \langle C, \{ \langle d_a,d_b\rangle  ,\langle c_a,c_b\rangle \},I\rangle$, then $disconnect^-$, read ``upwards" tells us that we must show how to construct a system of the form $A \vDash \langle C, \{ \langle d_a,d_b\rangle\},I\rangle$ and also show that $c_a, c_b \in C$. So, starting with a desired system, and using the rules upwards, we get some guidance, via pattern matching the system we want with the conclusions of the rules, as to how to build it. If we continue this process along all branches of the proof tree that we thus construct until we reach its tips, which require no further proof (for example the system $\emptyset \vDash \langle \emptyset, \emptyset,\emptyset \rangle$ is trivially constructible and usable; it is like the \emph{zero} of usable systems) then by reading the proof tree ``forwards" we see both how to construct our desired system and have a proof that it is usable. 

Our new logic (for usable systems) inherits the internal consistency of the underlying logic since the new logic was produced via conservative extension, and so it is sound, which means that any system constructed via the rules is guaranteed to be usable. 

\section{Introducing usability-enhancing components}

Next we can introduce \emph{usability-enhancing} (not merely ``interactive") components which some components in $I$ might be, or that can play the role of, \emph{wrappers} that \emph{guard} the rest of the system against non-usability to make $I$ components usable.  Further, we might have some abstraction mechanism which hides interactions (within the interactive component) which are deemed unusable by, say, the component's designer.

Then we have a condition that simplifies the structure by making the $A_i$ smaller by shrinking the family of sets $A$: if some components have been proved to be usability-enhancing, or been proved the protect the system against undesirable (and otherwise non-usable components---i.e. components that we might want in the system because of some very useful properties they have, but which are otherwise appallingly non-usable) by wrapping them up or filtering out or restricting their undesirable features, then we can take away some of the assumptions (which is what $A$ is essentially giving us) and get a simpler design.


As we said above, almost all components will be in almost all assumption sets, due to the requirements of the construction method, and its focus on guaranteeing that only usable systems are constructible. So, remember we had the idea of a contract of the form 
\[
A \vDash <C,N,I>
\]
where $A$ is to be thought of as an assumption  that {\it as long it  is acceptable that the components granted access to interactive elements are as given in $A$} then the system is usable. 

\begin{figure}
\centering
\includegraphics[width=10cm]{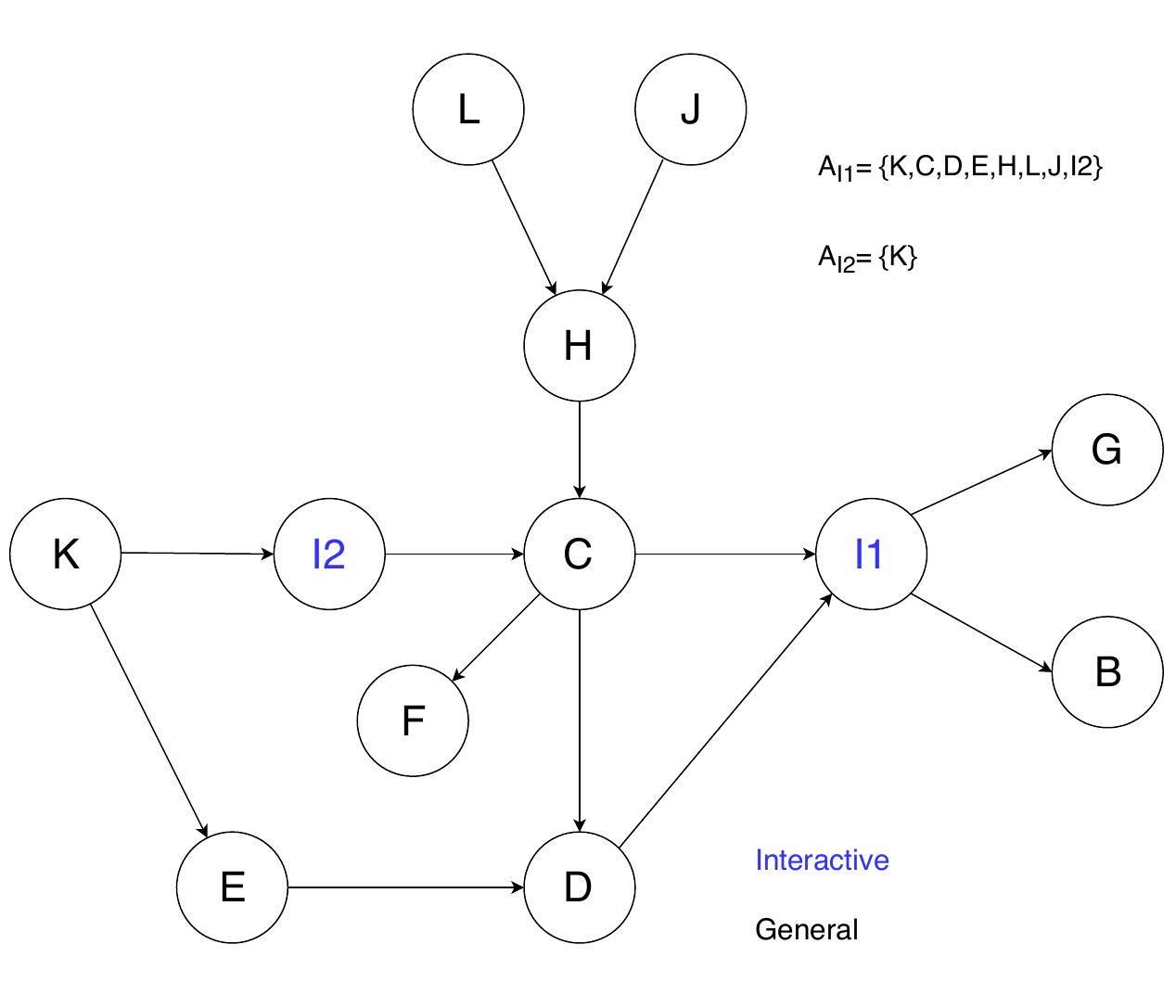}
\caption{A typical interactive system, without usability-enhancing components}
\label{fig:noTrust}
\end{figure}

Figure \ref{fig:noTrust} is a small example system built using the rules introduced in previous sections. We can see that the assumption sets contain almost all components. While we have a usable system, when those assumptions are taken into account, it would clearly be better (fewer assumptions means fewer opportunities for making an error or a wrong assumption about usability) to simplify the system in regard to the assumption sets, and this is where the idea of usability-enhancing components comes in. Specifically, given the size of the family $A$ in this case (composed from the union of the sets $A_{I_1}$ and $A_{I_1}$), the obvious question is: can we make it smaller (and hence make the system simpler)? This is where the notion of a usability-enhancing component enters the story.

A component is $usability-enhancing$ if we have verified that it can itself be assumed to be given access to any interactive component, and that it provides a barrier which stops components from accessing an interactive component inappropriately (i.e. maliciously or in some way detrimental to the system). Since it provides such a barrier, components which have to ``go through" the usability-enhancing component to access an  interactive component do not need to be in the set of components assumed to access that interactive component since we have verified that the usability-enhancing component stops any unusable behaviour. So, having designated a component as usability-enhancing, we can perform a simplification operation on the system in order to drop components from the assumption sets since we no longer need to make assumptions about them while preserving usability.

The idea of a usability-enhancing component forming a barrier is formalised by saying that if a usability-enhancing component $t$ exists between a component $n$ and an interactive component $i$, then that component $n$ does not need to be recorded as being in the set of components allowed access to that interactive component $i$. There is a proviso here: if the component $n$ we are considering can access the component $i$ via another path that does \emph{not} go through the usability-enhancing component $t$ then it clearly must stay in the assumption set for that interactive component.


In readiness for this we extend the idea of a system so that it now includes a set $U$ of usability-enhancing components (a subset of the set of all components). A system is now of the form $A \vDash <C,N,U,I>$. It is clear that if formerly we had the system $A \vDash <C,N,I>$ then we now have the equivalently usable system $A \vDash <C,N,\{ \},I>$, since there are no (as yet) usability-enhancing components in this system.

Then we have the following result, which codifies the reasoning above, including the restrictions.

\begin{lemma} Healthiness due to usability-enhancing components\\
Consider a system $\Sigma$ of the form $A \vDash <C,N,U,I>$. Assume one of the components $u$ in $C$ is designated as a $usability-enhancing$ component (as in the discussion above). Then we have an equivalently usable system $\Sigma'$ of the form $A' \vDash <C,N,U \cup \{u\},I>$ and $A'$ is related to $A$ as follows:

\begin{itemize}
    \item 
    for all interactive components $i \in I$, all components of $A$ that are {\bf only} on paths that are prefixes of paths from $u$ to $i$ are removed from $A_i$
\end{itemize}

The set of components that remains from $A_i$, due to the clauses above, form $A'_i$.
\qed
\end{lemma}

The {\bf only} in the clause above arises due to the proviso that a component may have another path to the  interactive component where the interactive component is not ``protected" by the trusted component, and we need to be aware of this and guard against trusting such a component.


%


\begin{figure}
\centering
\includegraphics[width=10cm]{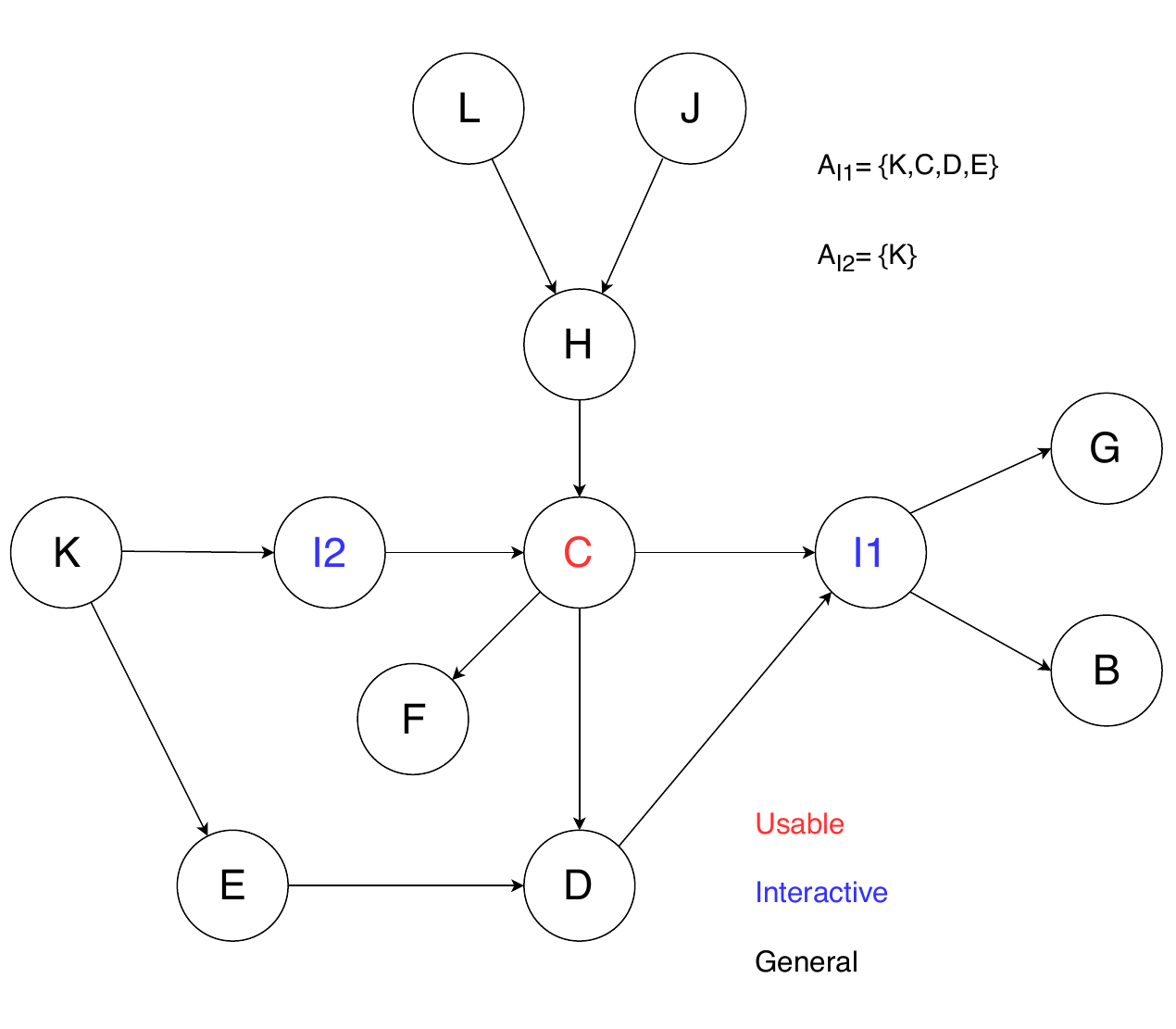}
\caption{Example system with one usability-enhancing component}
\end{figure}

\begin{figure}
\centering
\includegraphics[width=10cm]{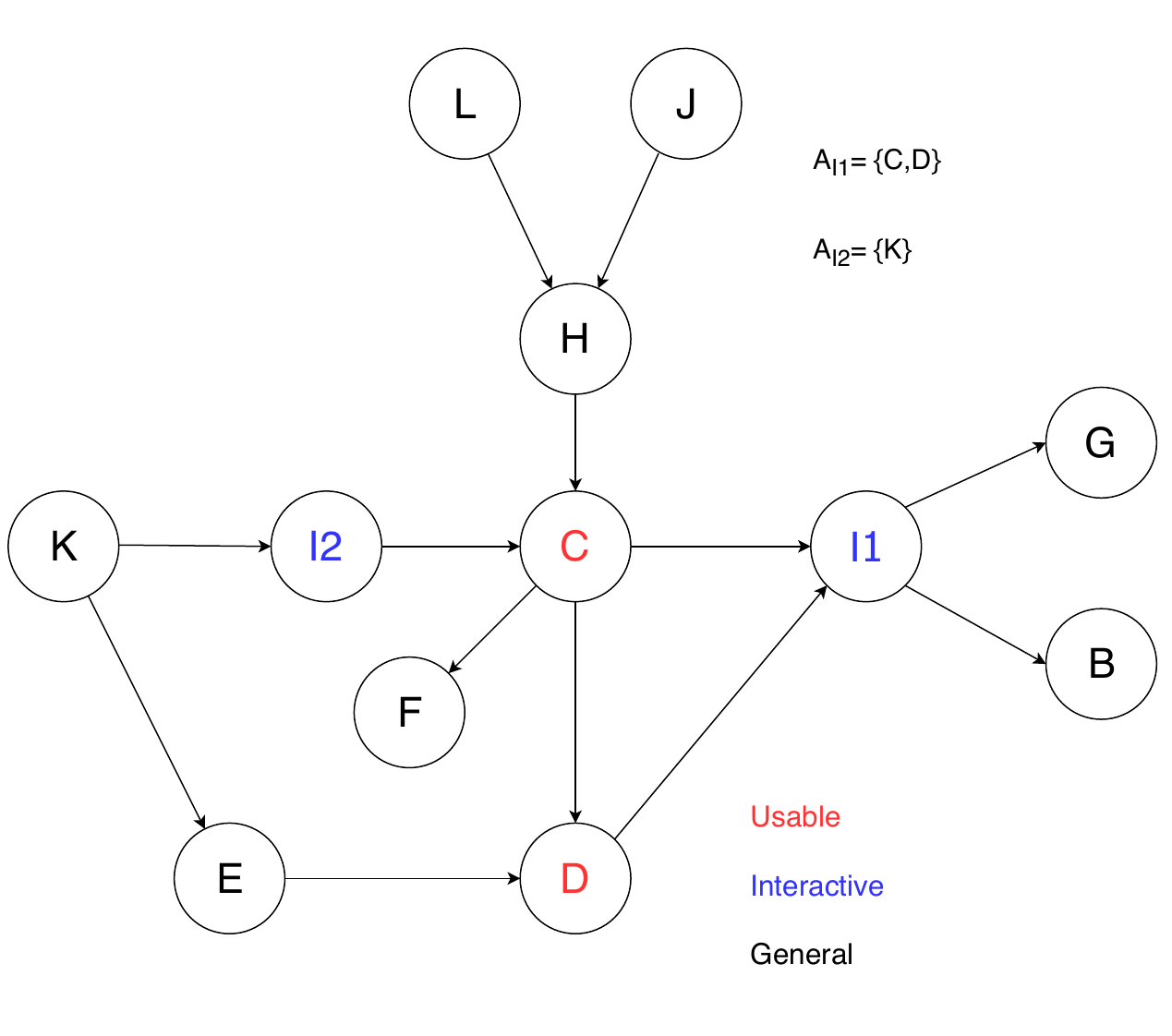}
\caption{Example system with two usability-enhancing components}
\end{figure}

\section{Discussion}

Recall that one of the aims of this paper was to give usable-by-construction rules for systems with usability-enhancing, interactive and general components in the context of designing, at an initial abstract level, usable systems 

This aim turned out to be very hard to achieve because of all the conditions around when it is usability-preserving to connect two components. We at first tried this using more conventional, \emph{ad hoc} means (some of this can be seen in a different guise in for example \cite{Rimba:2015}) but we could never convince ourselves, let alone prove, that the rules for building such a system were sound. This happened because the rules themselves had many provisos and hence reasoning about how they interacted when we used them to build large systems soon became impossible. Clearly, another approach was needed.

We used two strategies: start with a very few, simple (which means: no or few provisos) basic rules, and build more complex systems with a very few rules of composition (which take existing---probably partial---systems and combine or compose them into larger systems); do not try to ``do everything at once". This led to the idea of first building systems without considering notions of usabilty-enhancing components, and then, once such systems could be built, add a notion of such components and see how the existing systems without them could be transformed into systems with them, our final goal.

Now have a simple, stable basis upon which to build more sophisticated derived rules, if we need them. 

It is possible that we have paid a price for all this: perhaps few systems are actually now constructible with our simple rules. This might (depending on your point-of-view) be outweighed by being much more certain that the system we \emph{can} build are better structured and more amenable to reasoning. We, of course, believe that this is a price that is very well-worth paying. The simplicity and regularity of the rules means that we now have a hope of proving soundness and completeness, which was impossibly complex before. Ultimately we will be able to design provably usable systems.

\section{Conclusions}

The work recounted here, which centres around a need to consider the ways of designing and constructing a usable system made of various components, has several properties:

\begin{itemize}
\item
It allows an abstract characterisation of a usable system;
\item
It has logical rules that allow for checking and construction;
\item
Any complicated system may be built in terms of simpler ones using a small set of operations;
\item
We may use it as a basis for deriving further construction operations.
\end{itemize}

The rules here are, of course, tedious to use by hand (as the examples show), but we can express the rules very directly in the various proof assistants available (\eg\ PVS \cite{PVS92}) or perhaps program them in a language that already deals with search, like Prolog. Then by giving the desired system as conjecture to the proof assistant or a goal to a Prolog program, it can be used to (mainly automatically, given the simplicity of our rules) then construct a proof that the system is usable. For realistically large systems, with hundreds of components, this would be an important feature.

Another way of seeing this utility is to acknowledge that one the problems with realistically large systems is keeping track of dependencies (like our ``allowed access to interaction'' idea) and the rules given here do that. The fact that a large system, once built, can automatically be checked for conformance to requirements of dependencies is obviously valuable (even if the idea of having a logic to construct such system does not appeal).

There remains the question of how a very abstract model, once we have one, can be used as the basis for an implementation.  Our expectation would be to proceed via the existing and well-known and established techniques that are called \emph{refinement} \cite{DeB14}, which would take us from a design that provably has the required properties (\ie\ built with usability as a constructed and provably existing property) to provably usable implementations, since the central point of refinement is that it allows us to move from abstract to concrete (design to  implementation) while preserving meaning and properties. Taken together, then, we have rules that allow us to design and specify usable systems, and refinement rules that, preserving usability, take us to implementations or provably usable systems. 


Future work involves taking the concepts in ISO 9241 which seek to define usability and seeing how they interact with the work here.

Finally, this abstract system does not have the interpretation of components decided in any way, though the idea of ``interactive" does begin to impose one. However, we have a set of rules here which simply allows construction of connected components where \emph{some needed properties which make a system ``proper"} can be kept track of---so this has general application.

\section{Acknowledgments}
I worked with Paul Rimba on and off for a couple of years on the topic of ``secure-by-construction" which, it turns out, can be treated in ways very similar to those recounted here (basically replace ``usable" by ``secure" and ``usability-enhancing" by ``secret" in all the above). The current paper borrows heavily from work that we wrote up in draft form (but which is unpublished). I am grateful to Paul for the time and hospitality he showed during that work, to say nothing of the ideas and effort he put into it.

\bibliographystyle{plain}
\bibliography{refs}

\end{document}